\documentclass[conference]{IEEEtran}
\IEEEoverridecommandlockouts
\usepackage{cite}
\usepackage{amsmath,amssymb,amsfonts}
\usepackage{graphicx}
\usepackage{textcomp}
\usepackage{xcolor}
\usepackage{graphicx}
\usepackage{subcaption}
 \usepackage{booktabs} 
 \usepackage{pifont}   % for \ding symbols
\usepackage[utf8]{inputenc}
\usepackage{graphicx,subcaption}
 \usepackage[hidelinks]{hyperref} % in the preamble

 \usepackage{algorithm}          % the float
\usepackage[noend]{algpseudocode} % algorithmicx style

\newcommand{\cmark}{\ding{51}} % ✓
\newcommand{\xmark}{\ding{55}} % ✗

\def\BibTeX{{\rm B\kern-.05em{\sc i\kern-.025em b}\kern-.08em
    T\kern-.1667em\lower.7ex\hbox{E}\kern-.125emX}}
\begin{document}

\title{BIER-Star: Stateless Geographic Multicast for Scalable Satellite-Terrestrial Integration\\

}

\author{\IEEEauthorblockN{Mostafa Abdollahi,
		Wenjun Yang,
		Jianping Pan}
	\IEEEauthorblockA{University of Victoria, BC, Canada}

}

\maketitle

\begin{abstract}
The rapid expansion of LEO satellite constellations has enabled an integrated  terrestrial network and non-terrestrial network (TN-NTN), connecting diverse users such as aircraft, ships, and remote communities. These networks increasingly need a scalable and efficient multicast protocol for critical applications like emergency alerts, large-scale software updates, and real-time broadcasting. However, traditional multicast protocols, such as IP-based multicast and software-defined multicast approaches, introduce significant control overhead and struggle to adapt to the dynamic and mobile nature of satellite topologies. This paper presents BIER-Star, a stateless multicast protocol designed for the integrated TN-NTN. BIER-Star uses a two-layer geospatial gridding scheme (i.e., H3) to encode destinations as Earth- and space-cell identifiers rather than per-terminal addresses. This cell-based abstraction shortens the header bitstring, simplifies forwarding, and eliminates per-flow state and complex signaling. Our simulations indicate that BIER-Star reduces header size versus BIER and avoids geographic path-finding failures seen in greedy methods. 
\end{abstract}

\begin{IEEEkeywords}
Stateless Multicast, LEO Satellite Networks, BIER, H3 Gridding.
\end{IEEEkeywords}

\section{Introduction}
\par The integration of Terrestrial Network and Non-Terrestrial Network (TN-NTN) has significantly expanded global connectivity by extending internet access to regions beyond traditional infrastructure. This advancement has enabled a diverse array of users, including airplanes, ships, and remote communities, to benefit from satellite-ground communication. Consequently, new applications have emerged, such as satellite-based IPTV, emergency communication systems like Apple’s Emergency SOS via Satellite \cite{P10} and Iridium’s GMDSS \cite{P11}, and innovative services like Starlink’s Direct-to-Cell \cite{P12} and AST SpaceMobile, aimed at providing connectivity directly to smartphones in areas without terrestrial coverage. Despite this rapid growth in satellite services, highlighted by Starlink reaching over 7 million users by 2025, the prevalent reliance on unicast communication for tasks such as software updates remains problematic. Unicast results in substantial data duplication and bandwidth inefficiency, highlighting the need for a multicast protocol tailored to the unique characteristics of the integrated TN-NTN.

\par Multicasting in the integrated TN-NTN is challenging because satellites move rapidly, causing frequent handovers and link changes. IP multicast (e.g., PIM-SM/SSM \cite{P18}) depends on distribution trees and per-flow state (i.e., router records for each stream), making repeated join/prune updates impractical at the scale of thousands of satellites and millions of users.  Recent Low-Earth-Orbit (LEO) satellite multicast methods \cite{P1,P2,P3,P4,P5} require a centralized controller to compute routes, update satellites, and manage user terminal membership (including joins and leaves), which is costly at the scale of current constellations and traffic. By contrast, Bit Indexed Explicit Replication (BIER) \cite{P7} is a stateless protocol for terrestrial networks. It encodes the destination set as a bitstring in the packet header, and forwarders check that bitstring against per-interface bitmasks. However, in satellite networks BIER becomes impractical. The bitstring can grow to thousands or millions of bits, and frequent bitmask updates in dynamic topologies impose excessive overhead. 

\par
\begin{figure}[htbp]
    \centering
    \includegraphics[width=0.5\textwidth]{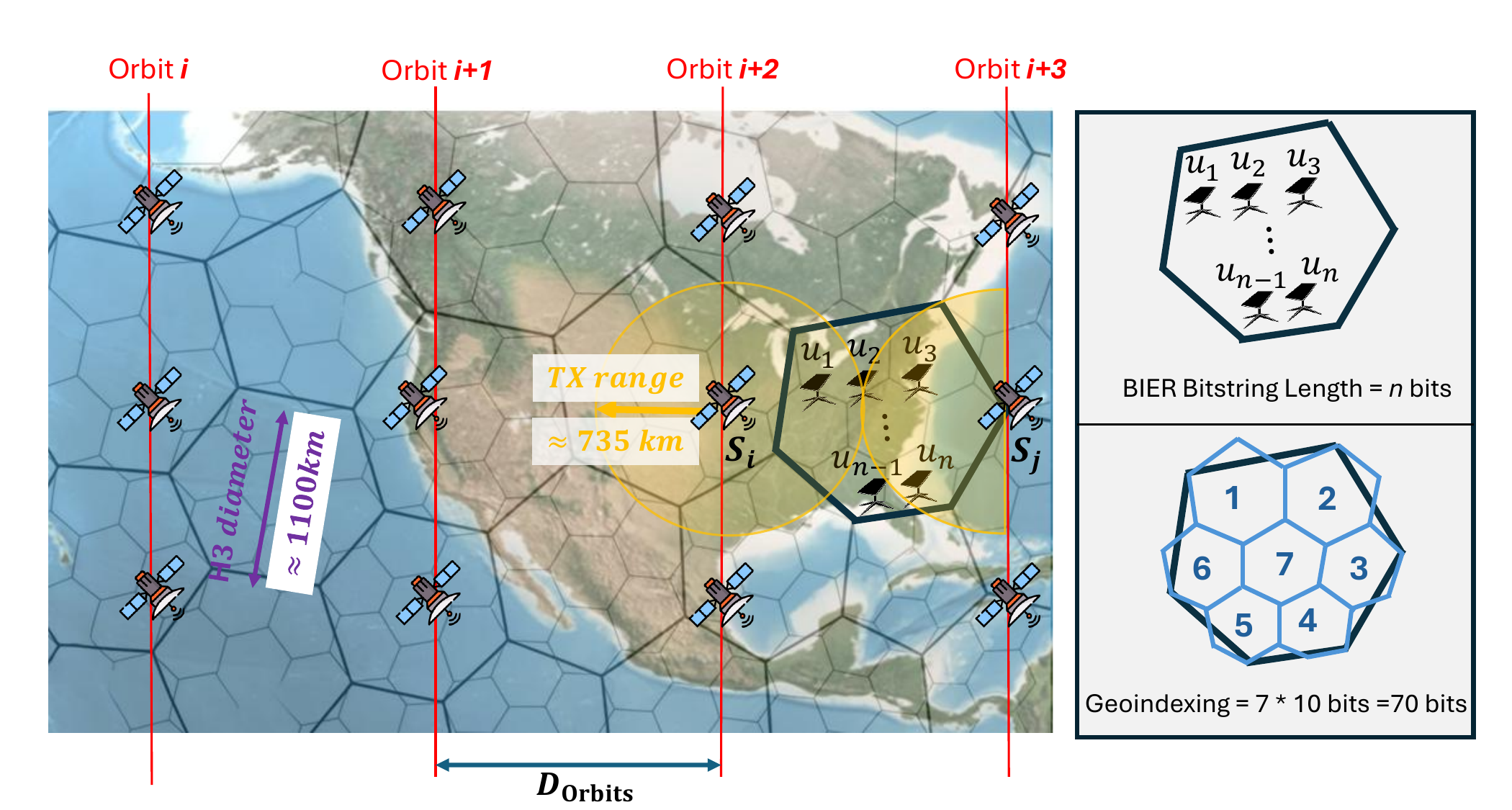}
    \caption{OneWeb's near-polar Walker Star topology causes orbital plane spacing ($D_{\text{Orbits}}$) to vary greatly, from about 3,900 km at the equator to minimal near the poles, resulting in a non-uniform satellite distribution within an H3 grid. }
    \label{fig:oneweb}
\end{figure}

 BIER efficiency in TN-NTN can be improved by exploiting the fixed geographic locations of user terminals and satellites so that addressing is region-based. In the standard scheme, one header bit is assigned per terminal \(u_i\), which is not scalable for large groups. As shown in Fig.~\ref{fig:oneweb}, instead of allocating \(n\) bits for \(n\) users, the regions containing the users can be addressed via geoindexing, reducing the header to tens of bits (e.g., 70 bits for thousands of users). However, adopting geoindexing introduces two issues that must be resolved: (i) a many-to-many satellite--cell mapping---multiple cells may be served by one satellite, and a single cell may be covered by multiple satellites (e.g., $S_i$ and $S_j$)---complicating the use of a single geoindex for both satellites and user terminals; and (ii) selection of an appropriate geoindex resolution for routing, since satellites may need to be represented at different cell sizes, as illustrated in Fig.~\ref{fig:oneweb}.
 
\par In response to these challenges, this paper introduces BIER-Star, an extension of the BIER protocol for the integrated TN-NTN. BIER-Star operates in two layers, the user layer and the satellite layer, by partitioning the earth and satellite network into H3 geographical cells. In the user layer, user terminals use the Internet Group Management Protocol (IGMP) \cite{P19} to send join/leave messages to their serving satellite in the satellite layer. The satellite registers the multicast group membership and refreshes it at each handover. Meanwhile, the source (a gateway or satellite) periodically computes multicast shortest path trees and encodes the corresponding cell IDs into the packet header. Forwarding satellites decode the header and forward packets to neighboring satellites whose cells match the next entries. Simulation results show that, by leveraging IGMP, BIER-Star significantly reduces bitstring length, and through its cell-based forwarding mechanism, we achieve comparable short delays without the overhead of maintaining per-flow multicast state. 

\par The main contributions of this paper are:
\begin{itemize}
    \item We propose a two-layer architecture (i.e., user and satellite layers) that leverages geo-indexed cell partitioning (i.e., H3) to map user terminals and satellites to geographic regions, which enables efficient addressing for routing and forwarding purposes.

\item We aggregate user terminals at the user layer and manage group membership via IGMP at serving satellites, which significantly reduces the packet header bitstring length compared to traditional BIER, while ensuring seamless updates during frequent handovers.

\item We propose a cell-based geographic routing method built on an SPT. Simulation results demonstrate that the proposed method is more reliable at finding paths than state-of-the-art solutions. 

\end{itemize}

\par In Sec. II, we present a brief overview of related works. Sec. III discusses problem definition and challenges. Sec. IV details the BIER-Star framework, while Sec. V analyzes BIER-Star in terms of bitstring length, satellite capacity, link failure resilience, and geographical routing. Finally, Sec. VI concludes the paper.

\section{Related Works}
We structure the related work as follows: multicasting in LEO satellite networks, and stateless multicasting.
\par {\bf Multicasting in LEO satellites:} To address scalability, a centralized-controller SDN multicast framework constructs multicast trees using global topology knowledge \cite{P1}. In addition, to improve robustness, a local subtree-reconstruction mechanism repairs failed segments without full recomputation \cite{P2}. It reduces segment routing overhead via source routing and centralized tree computation, minimizing forwarding state and onboard complexity. Authors in \cite{P3} construct QoS-aware multicast trees using a Weighted Rectilinear Steiner Tree (WRST) formulation to balance bandwidth savings with QoS requirements. Consequently, any changes in group membership, such as adding or removing user terminals, require the multicast tree to be updated, which creates overhead for the controller. The work by \cite{P5} adds adaptive group management and an obstacle-aware rectilinear Steiner tree that bypasses failed or congested links. However, frequent recomputation and dissemination of segment/replication state across a constellation of thousands of fast-moving satellites incurs high control-plane overhead. Additionally, a partitioning of the Earth into hexagonal blocks is proposed \cite{P16}. For each multicast flow, destination user terminals broadcast join/leave messages to all forwarding satellites so that the constellation learns the membership. Each forwarder then builds a local block tree from the destination set, and data packets—carrying a compressed destination list in the header—are forwarded geographically, hop by hop. However, this approach may be impractical in the presence of thousands of satellites and multicast streams, because each satellite must build a distinct block tree per stream. Additionally, the geographic hop-by-hop approach can suffer delivery failures due to the limited number of inter-satellite paths/links.

\par {\bf Stateless Multicasting:} In the stateless multicast architecture BIER \cite{P7}, the ingress router sets a bitstring in the packet header, where each bit corresponds to a unique egress router. At each hop, a forwarding router performs a bitwise AND between the packet’s bitstring and an interface’s bitmask to decide whether to replicate and forward the packet. Additionally, BIER-TE \cite{P6} maps bits to links rather than egress routers, while BIER-FRR \cite{P8} precomputes backup entries for local rerouting around link or node failures. Similarly, Yeti \cite{P9} uses four specialized label types to build a dynamic label stack that shrinks as packets traverse the network, with routers removing labels that are no longer needed. Unfortunately, these stateless schemes require one header bit per egress router or per link. In TN-NTN, this would inflate headers to millions of bits if a bit were allocated per user terminal or link. Additionally, every path change requires a centralized controller to update the satellites, which is challenging in highly dynamic networks.

\par Therefore, a scalable, stateless multicast architecture is needed for the integrated TN-NTN with thousands of satellites and millions of users.

\section{Problem Definition and Challenges}
\par
\begin{figure}[!t]
  \centering
  % 0.95\columnwidth = overall target width (change as needed)
  \resizebox{0.60\columnwidth}{!}{%
    \begin{minipage}{\columnwidth}
      \centering
      \begin{subfigure}[t]{0.48\linewidth}
        \includegraphics[width=\linewidth]{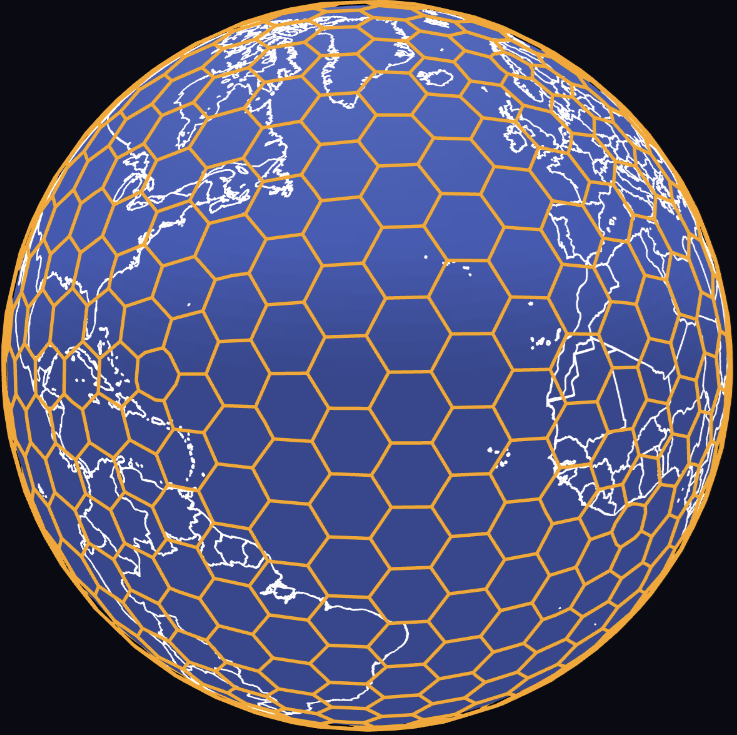}\caption{H3}
      \end{subfigure}\hfill
      \begin{subfigure}[t]{0.48\linewidth}
        \includegraphics[width=\linewidth]{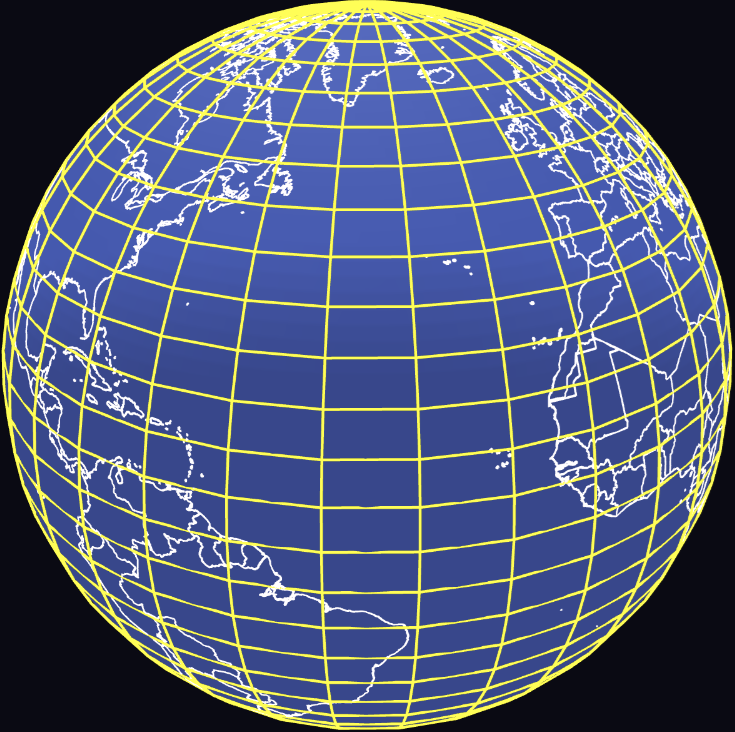}\caption{Geohash}
      \end{subfigure}

      \vspace{0.5ex}

      \begin{subfigure}[t]{0.48\linewidth}
        \includegraphics[width=\linewidth]{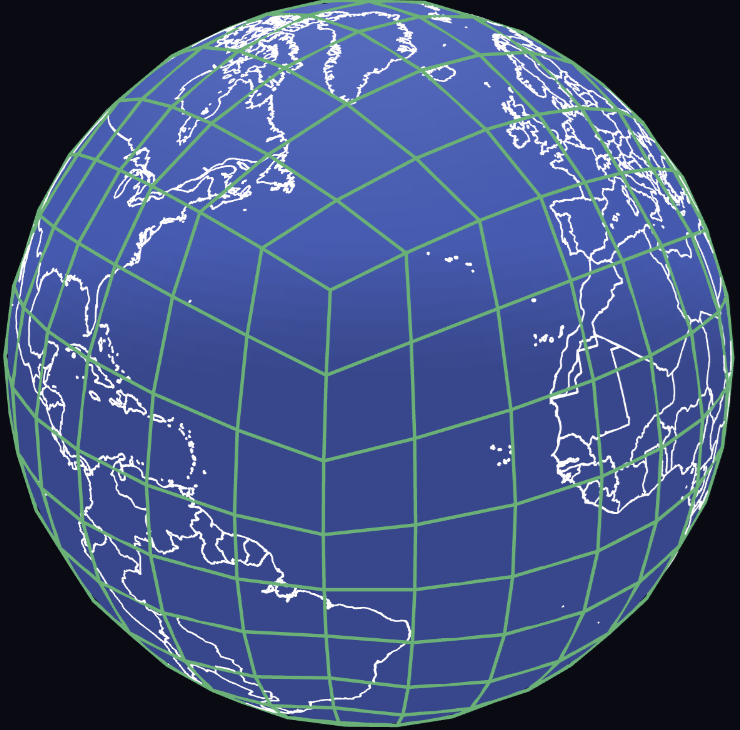}\caption{S2}
      \end{subfigure}\hfill
      \begin{subfigure}[t]{0.48\linewidth}
        \includegraphics[width=\linewidth]{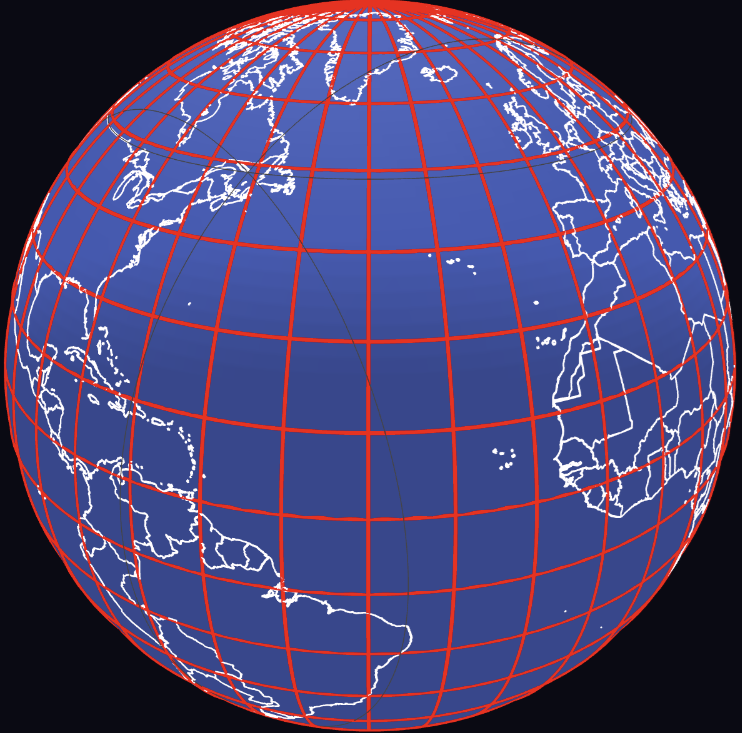}\caption{Latitude–longitude}
      \end{subfigure}
    \end{minipage}%
  }
  \caption{Gridding mechanisms.}
  \label{fig:Sec2-1}
\end{figure}

 In large-scale TN-NTNs with thousands of satellites and millions of user terminals, one-bit-per-egress schemes (as in BIER) yield headers of thousands to millions of bits, which are impractical. A more scalable approach is to aggregate receivers into stable geographic regions using a geo-indexed grid. Candidates include H3, S2, and Geohash, each with different trade-offs in cell uniformity, hierarchical aggregation, lookup cost, and encoding compactness (Fig.~\ref{fig:Sec2-1}). The first question is: {\it which gridding method is most suitable for the integrated TN-NTN?}

\par Moreover, a cell (region) can be served by more than one satellite; therefore, multiple satellites covering the cell must be managed for routing. However, applying a single grid to both ground users and satellites is difficult. As shown in Fig.~\ref{fig:oneweb}, a LEO satellite’s coverage area can overlap with multiple grid cells, and a single cell can be covered by multiple satellites. This creates a complex many-to-many mapping between satellites and cells that complicates data forwarding. Therefore, the second question is: {\it how should gridding be designed for user terminals and satellites?} Additionally, {\it what gridding resolution should be used based on the distributions of user terminals and satellites?}

\par In the next section, we will present BIER-Star, a stateless multicast approach that addresses these questions.

 \begin{figure}[htbp]
    \centering
    \includegraphics[width=0.48\textwidth]{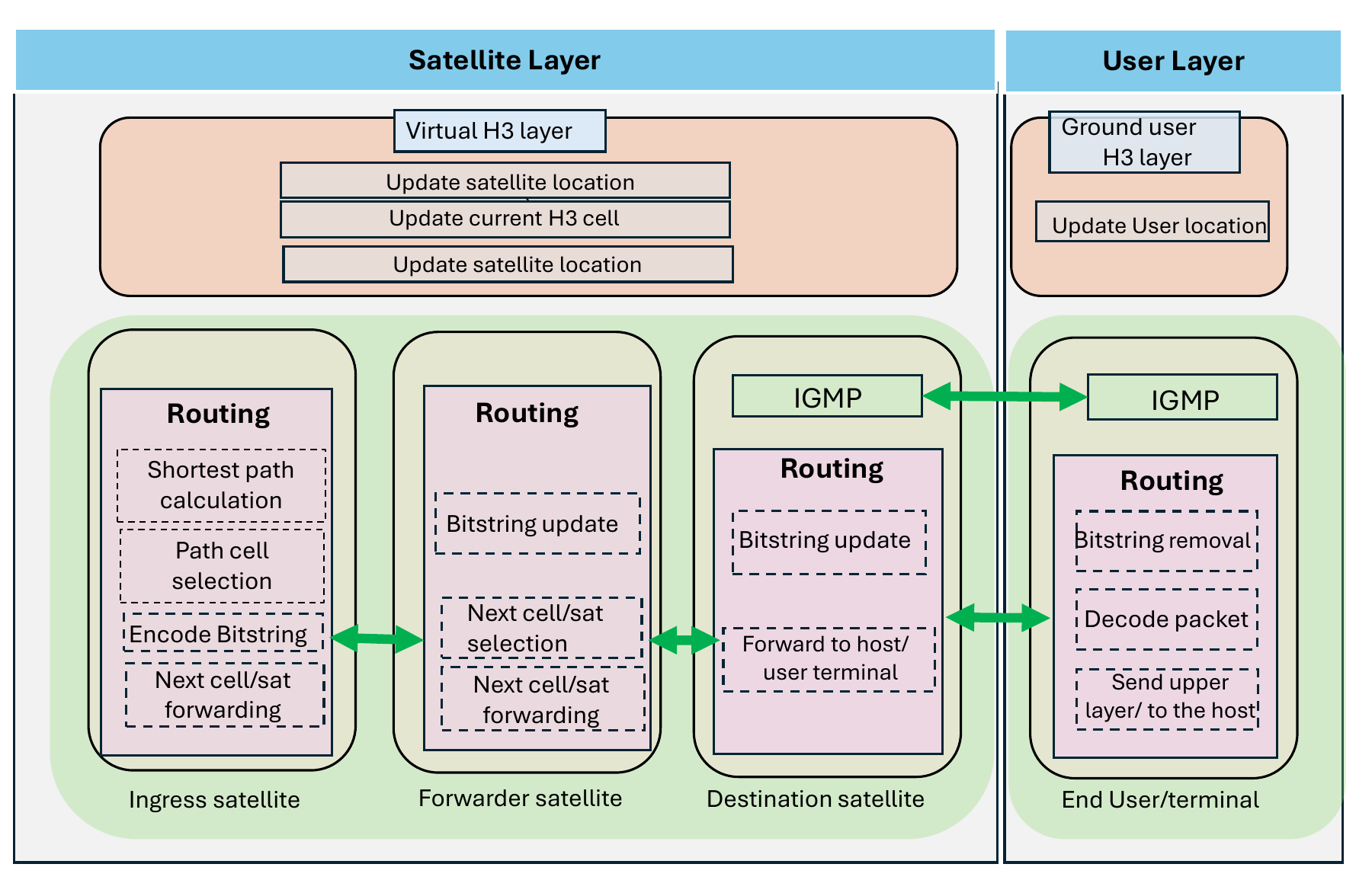}
    \caption{BIER-Star framework}
    \label{fig:framework}
\end{figure}

\section{BIER-Star}
\par BIER-Star (Fig.~\ref{fig:framework}) is built on a two-layer geographic abstraction with H3 cells, local IGMP, and stateless forwarding. In the user layer, terminals report their location and current H3 cell at each handover and join/leave groups via IGMP at the serving satellite. In the satellite layer, gateways/satellites maintain a virtual H3 map and neighbor-to-cell reachability and execute stateless packet encoding/decoding and next-hop selection, replicating at branch points. In the following, we discuss why H3 geoindexing method is better than others, then satellite and user layers, membership management, routing/forwarding, and packet encoding are discussed. 

\subsection{Why H3 Geoindexing}
Geographical gridding methods, including H3 \cite{P20}, S2 \cite{P21}, Geohash, and latitude/longitude, partition Earth into discrete, addressable cells (see Fig.~\ref{fig:Sec2-1}). As summarized in Table~\ref{Table1}, H3 is well suited to multicast routing: its hexagonal cells maintain near-uniform area globally, avoiding the polar distortion of latitude/longitude and Geohash, and they offer predictable adjacency (typically six neighbors). Compared with square/rectangular grids such as S2, H3 enables more balanced neighbor analysis and simpler local lookups \cite{P20}. This combination of uniform coverage and stable adjacency makes H3 a robust, predictable substrate for complex spatial operations.  

\par
\begin{figure}[!t]
  \centering

  \begin{subcaptionbox}{Starlink PoPs and Gateways\label{fig:png}}[\columnwidth]
    {\includegraphics[width=0.6\columnwidth]{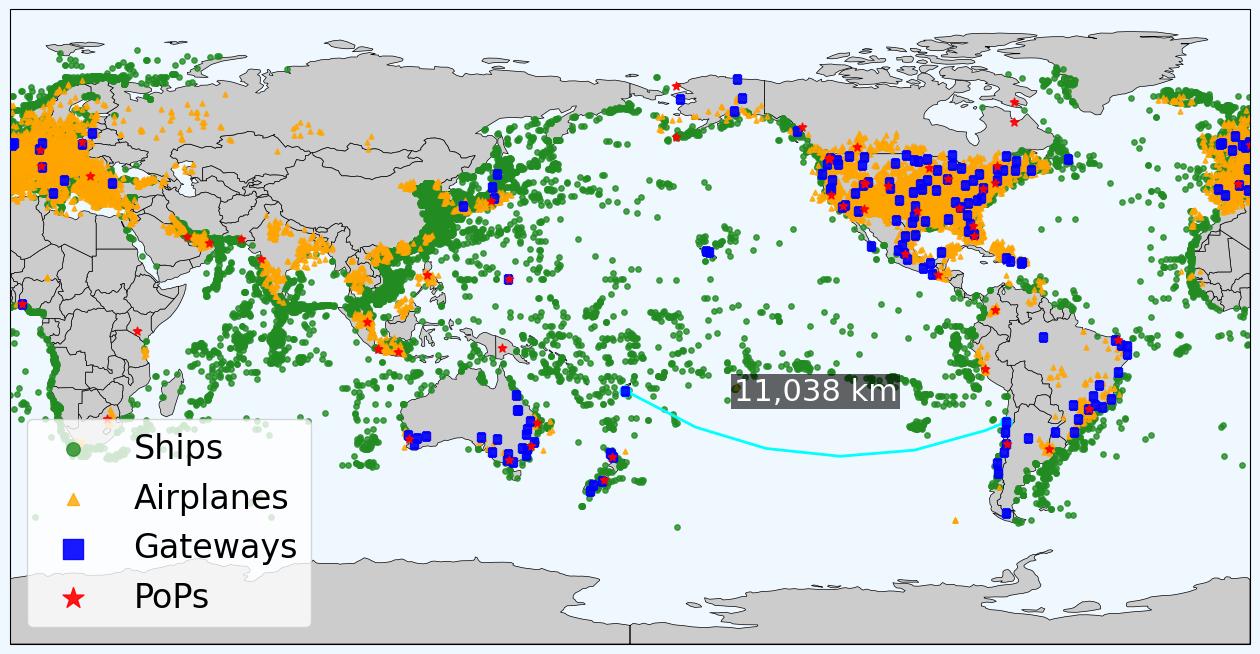}}
  \end{subcaptionbox}

  \par\medskip

  \begin{subcaptionbox}{Required bits per cell\label{fig:eps}}[\columnwidth]
    {\includegraphics[width=0.6\columnwidth]{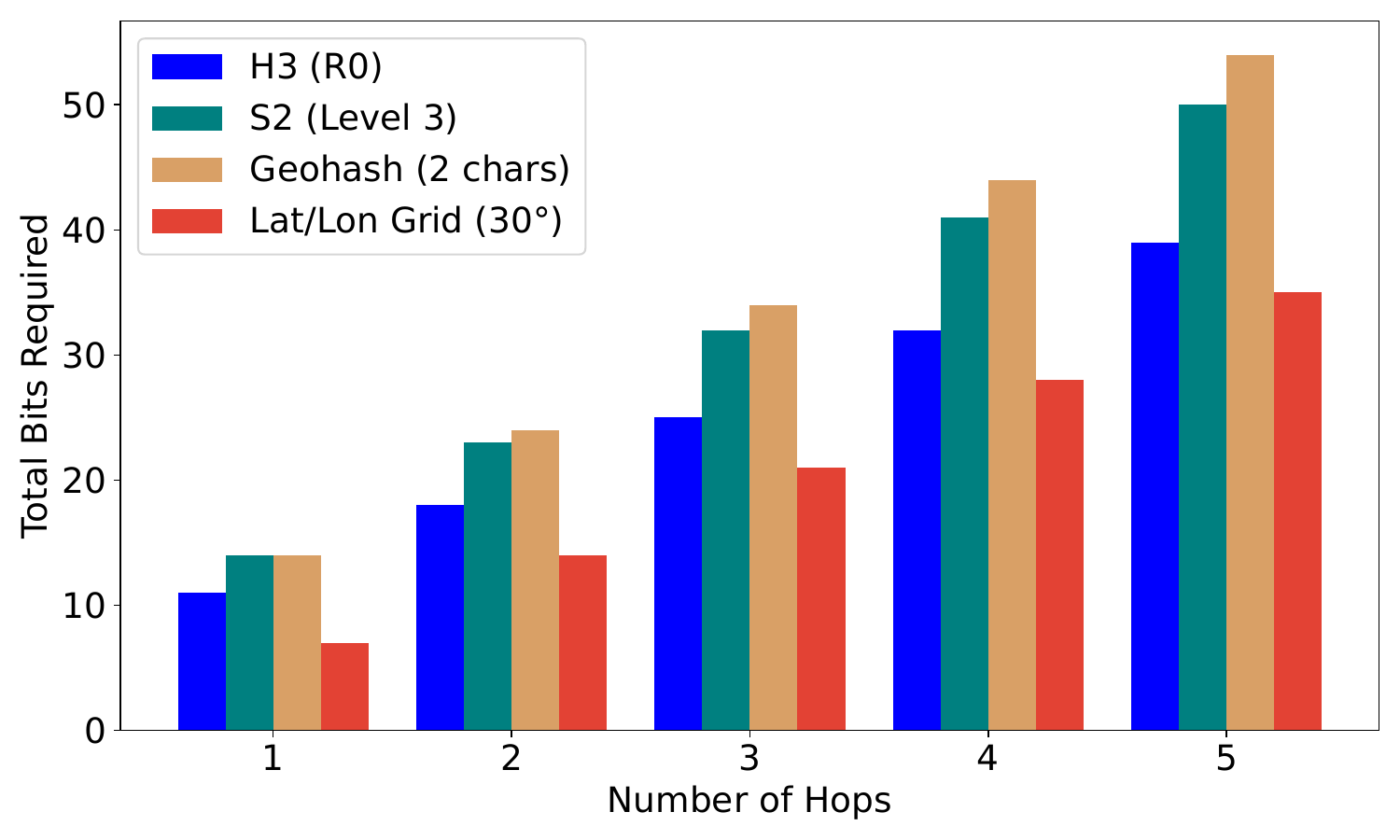}}
  \end{subcaptionbox}

  \caption{(a) Starlink PoPs and Gateways; (b) number of required bits per cell in different geo-indexing methods.}
  \label{fig:stacked}
\end{figure}

 With more than 7{,}900 LEO Starlink satellites, a distributed network of ground stations (Fig.~\ref{fig:stacked}), and an assumed maximum satellite--gateway slant range of $\approx$940\,$\mathrm{km}$, end-to-end paths for South Pacific air/sea routes can span $\approx$5{,}500\,$\mathrm{km}$—about five satellite hops. We compare the bits needed to encode grid identifiers across geo-indexing schemes (H3 at resolution 0, S2 at level 3, Geohash with two characters, and a $30^\circ$ latitude/longitude grid), assuming a few bits indicate the scheme/resolution and additional bits encode the cell per hop for paths of 1--5 hops. Under comparable granularity, H3's uniform tiling yields shorter encodings than S2 and Geohash. 

\begin{table}[t]
\caption{Comparison of H3, S2, Geohash, and latitude-longitude grid for multicast-relevant criteria.}
\centering
\label{Table1}
\renewcommand{\arraystretch}{1.4}
\begin{tabular}{lcccc}
\toprule
\textbf{Criteria} & \textbf{H3} & \textbf{S2} & \textbf{Geohash} & \textbf{Lat-Lon} \\
\midrule
Bit length compactness     & \cmark & \cmark & \xmark & \xmark \\
Neighbor lookup efficiency & \cmark & \xmark & \xmark & \cmark \\
Multicast routing suitability & \cmark & \cmark & \xmark & \xmark \\
Uniformity (cell area)     & \cmark & \cmark & \xmark & \xmark \\
Polar region suitability   & \cmark & \cmark & \xmark & \xmark \\
\bottomrule
\end{tabular}

\end{table}

\subsection{BIER-Star Framework}

\par {\bf Satellite and User Layers:} In Fig.~\ref{fig:framework}, we propose a dual-layer architecture over H3, named BIER-Star, in which a user layer that aggregates membership and a satellite layer that handles routing/forwarding. The layers maintain synchronized H3 namespaces but may use different resolutions (e.g., coarser in space, and finer on ground). A ground H3 cell is typically covered by multiple satellites, and each satellite footprint overlaps with many ground cells. However, each terminal attaches to exactly one serving satellite at any given time. This separation confines all group management to the user layer, allowing the satellite layer to remain \emph{stateless}. The satellite layer can then forward traffic using a virtual H3 map and neighbor-to-cell reachability, all without storing user-specific state. As a result, this design reduces network overhead, prevents satellites from having to track every data stream, and ensures reliable communication even during rapid handovers.

\par {\bf Membership management:} The BIER-Star architecture manages multicast group membership using a combination of localized IGMP signaling and end-to-end control updates. Initially, user terminals that want to receive a multicast flow send a request to the ingress router. In addition, a user terminal initiates membership by sending IGMP Join or Leave messages directly to its currently connected satellite. This localized exchange enables the satellite to maintain awareness of which terminals are interested in specific multicast flows. In addition, during every handover, user terminals send an IGMP Join message to the new satellite that they connect to. Crucially, these IGMP messages are confined to the local satellite link and are not propagated to ingress routers or multicast sources. To remain active, user terminals periodically send membership updates directly to the multicast flow’s source or ingress router, confirming their interest and identifying their current serving satellite. If an update is not received within a predefined interval, the ingress router presumes the user is inactive and removes the associated satellite’s H3 cell from the multicast bitstring. {\it With this membership mechanism, user management is confined locally, in contrast to stateless protocols \cite{P7, P8, P6, P9} that manage egress routers in the packet header and the approach in \cite{P16} that informs all satellites about user terminals—a process that is extremely costly with thousands or millions of users.}

\begin{algorithm}
\caption{BIER-Star Encoding Method}
\label{alg:bierstar-static}
\begin{algorithmic}[1]
    \Statex \textbf{Input:} \quad Satellite graph $G=(V,E,W)$; source satellite $s_{\text{src}}\in V$; \quad User terminals $U=\{u_j\}_{j=1}^M$ with locations $\mathrm{loc}(u_j)$; Resolution $r$.
    \Statex \textbf{Output:} \quad Encoded packet header $\mathcal{H}$.
      \State $D \gets \emptyset$ // Identify destination satellites for all users.
        \For{$j \gets 1$ \textbf{to} $M$}
            \State $d_j \gets \mathrm{assign}(\mathrm{loc}(u_j))$
            \State $D \gets D \cup \{d_j\}$
        \EndFor
        \State // Find the shortest path tree from the source satellite.
        \State $\mathrm{par} \gets \text{Calculate\_Shortest\_Paths}(G, s_{\text{src}})$ \Comment{Returns parent pointers}

        \State /* {\it Create geographic "footprints" of the paths for the given resolution.} */
        \State Initialize map of sets $C_{i} \gets \emptyset$.
        \For{each destination $d \in D$}
            \State $\mathrm{path} \gets \text{Reconstruct\_Path}(s_{\text{src}}, d, \mathrm{par})$
            \For{each satellite $v \in \mathrm{path}$}
                \State $\mathrm{pos}_v \gets \mathrm{get\_position}(v)$
                \State $C_{i} \gets C_{i} \cup \{\mathrm{H3}(\mathrm{pos}_v, r)\}$
            \EndFor
        \EndFor
               
        \State //  Encode the footprints into a packet header.
\State \resizebox{\linewidth}{!}{$
  \mathcal{H} \gets \mathrm{Assemble~}\!\big(\text{version},\ \{\, (i, \mathrm{Enc}(C_i)) \mid i \in \mathrm{sorted}(\mathrm{keys}(C)) \,\}\big)
$}

\end{algorithmic}
\end{algorithm}

\par
\begin{figure}
    \centering
    \includegraphics[width=0.40\textwidth]{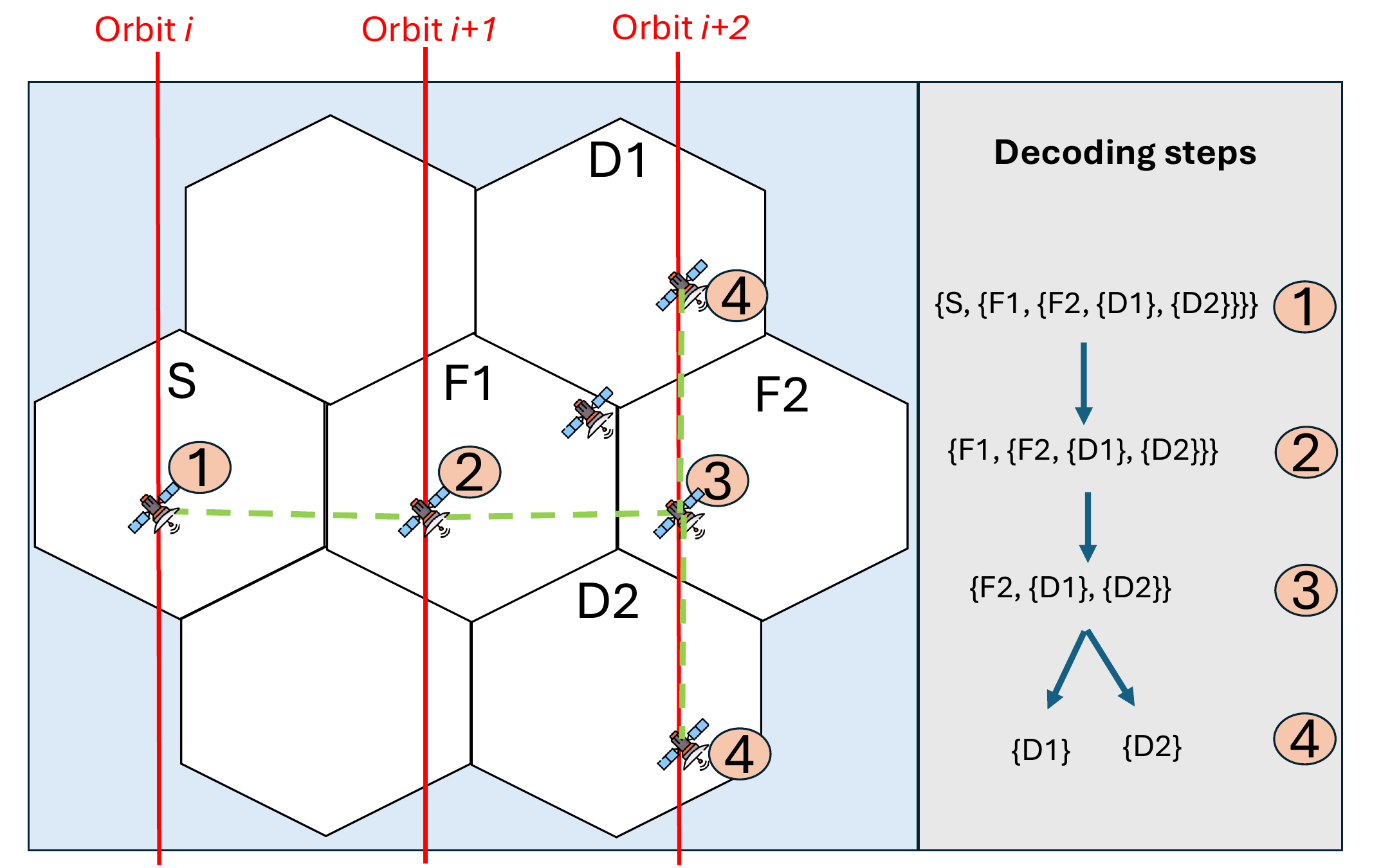}
    \caption{Decoding process in BIER-Star.}
    \label{fig:BIER-Star-Decoding}
\end{figure}

 {\bf Routing: }BIER-Star's routing exploits predictable handover epochs in LEO (e.g., $\approx$15s handovers \cite{P22}, and $\approx$2--3 min visibility in Starlink). In Algorithm~\ref{alg:bierstar-static}, the ingress maps each user's location to its serving satellite under a selection policy (e.g., nearest) and collects the destination set (lines 1--4). It then computes a single-source SPT on the ISL graph from the ingress to all destination satellites (line 6), reconstructs each path, and---at the chosen resolution $r$---collects the H3 cell footprints that the paths traverse (lines 8--13). Finally, it serializes the per-shell cell sets into the packet header (line 15), with a version and resolution tag per shell block. As shown in Fig.~\ref{fig:BIER-Star-Decoding}, the gateway encodes the cell path  \({S}\!\to\!{F1}\!\to\!{F2}\!\to\!\{{D1},\,{D2}\}\) into the packet header. After \({S}\) and \({F1}\) decode and forward the packet, satellite \({F2}\) replicates it and sends copies toward \({D1}\) and \({D2}\).

\par Moreover, in typical LEO shells, each satellite connects via ISLs to a small, fixed set of neighbors (often two in-plane and two cross-plane), and these links—and their handovers—follow a predictable schedule. Each epoch, the satellite builds a small target-cell routing table that maps H3 target cells to the next hop that makes geographic progress, with optional 2–3-hop backups for fast reroute. Upon receipt, the satellite reads the next target cell in the header and forwards to the mapped next hop; if that link is down, it chooses a backup that still advances toward the cell. This removes per-flow state from satellites, aligns forwarding with geography, and stays robust under frequent, scheduled handovers.

\section{Evaluation}

\par We evaluate BIER-Star on Starlink and OneWeb in the StarPerf simulator using two topology sources: a “+grid” synthetic topology for resilience, routing, and bitstring analyses, and TLE-driven snapshots for dwelling time analysis. For user terminals, we use an aircraft dataset collected over 100 time slots across one week from flightradar24 website\footnote{\url{https://www.flightradar24.com/}}. We used the following performance metrics: 
\begin{itemize}
    \item {\bf Bitstring (header overhead)}: the total number of header bits required to address all destinations within the region (cell/partition) containing the maximum number of user terminals (i.e., the worst-case region).
    
    \item {\bf Average reach rate (routing reliability):} the fraction of target destinations for which the routing method finds a valid path from the source (i.e., reachable destinations over total destinations). 

\item {\bf Dwelling time:} The duration a satellite–cell association remains within the same H3 cell at resolution $r$. We approximate it by projecting orbital motion onto the ground track (ignoring Earth’s rotation), with ground–track speed:
\begin{equation}
v_g=\frac{R_E+h}{R_E}\,v_o\cos i,
\label{eq:vg}
\end{equation}
where $R_E$ is Earth’s radius, $h$ the satellite altitude, $v_o\approx 7.66~\mathrm{km/s}$ the orbital speed in LEO, and $i$ the inclination (radians). For an H3 cell of effective diameter $d_r$, the dwelling time is:
\begin{equation}
T_{\text{cell}}(i,h,r)\approx \frac{d_r}{v_g}
=\frac{R_E}{R_E+h}\cdot\frac{d_r}{v_o\cos i}.
\label{eq:tcell}
\end{equation}
Thus, larger cells (lower $r$) increase dwelling time, and higher inclination (smaller $\cos i$) reduce $v_g$ and therefore increase dwelling time.

\item {\bf Link/Node removal (resilience):} the largest number (or fraction) of ISLs (links) or satellites (nodes) that can be removed within the selected H3 cell set at resolution $r$ without disconnecting the source from the destination set.

\end{itemize}

\begin{figure}[!t]
  \centering
  % \captionsetup[sub]{justification=centering,singlelinecheck=false}

  \begin{subfigure}[t]{\linewidth}
    \centering
    \includegraphics[width=0.7\linewidth]{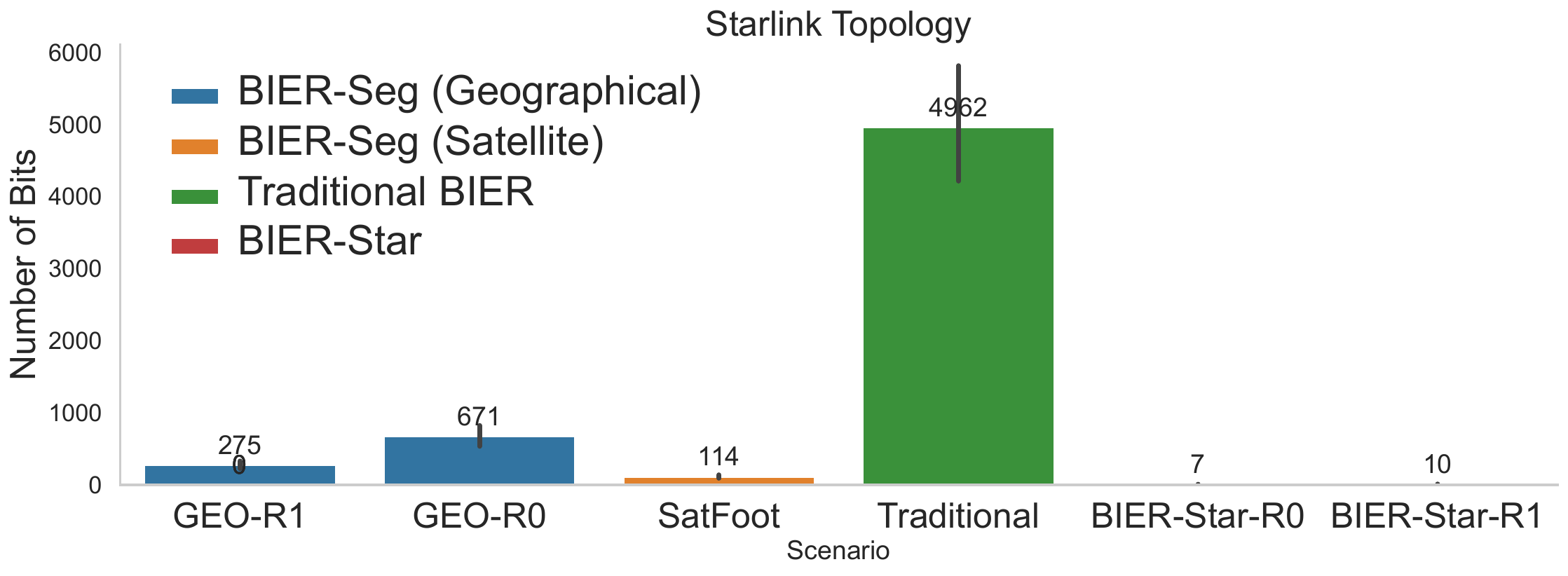}
    \caption{Starlink}
    \label{fig:starlink}
  \end{subfigure}

  \vspace{0.7em}

  \begin{subfigure}[t]{\linewidth}
    \centering
    \includegraphics[width=0.7\linewidth]{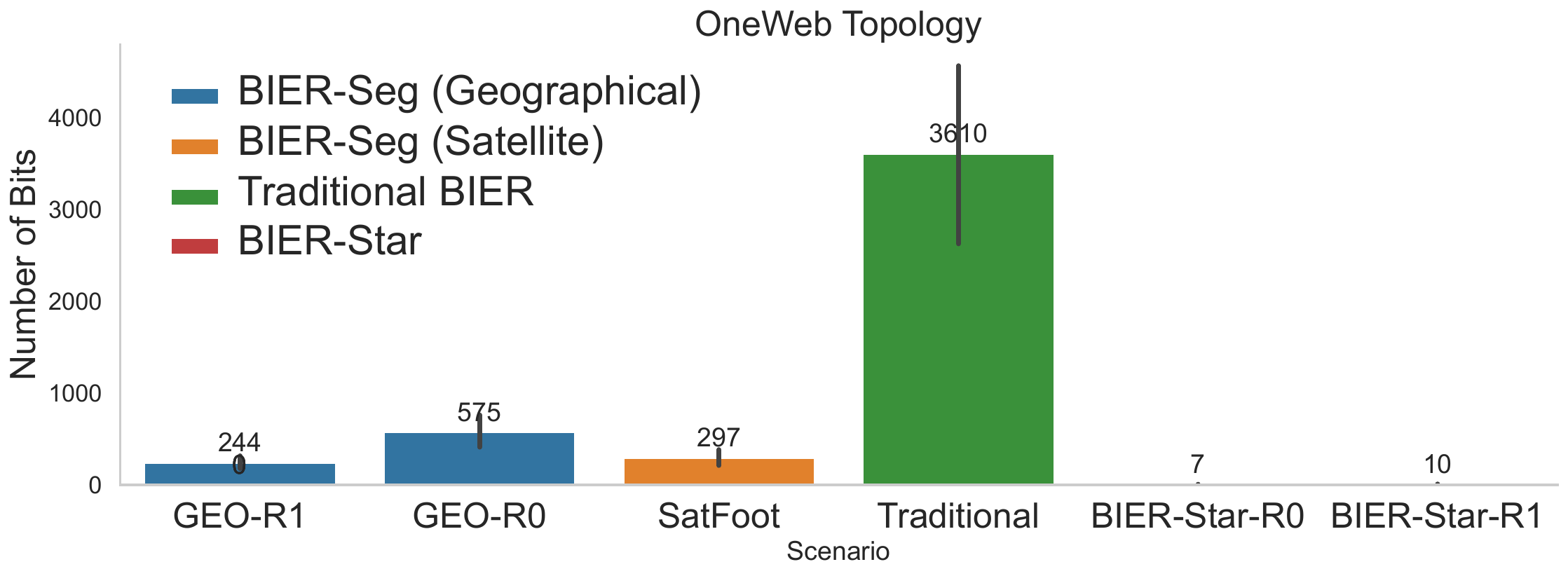}
    \caption{OneWeb}
    \label{fig:oneweb2}
  \end{subfigure}

  \caption{Bitstring comparison}
  \label{fig:bitstringcomp}
\end{figure}

\par Fig.~\ref{fig:bitstringcomp} compares the maximum per-cell bitstring length for OneWeb and Starlink with airplanes as user terminals. We evaluate (i) BIER-Star; (ii) classic BIER (“Traditional”), which assigns one bit per user terminal; and (iii) segmented-BIER variants that partition the network by H3 cells at resolutions 0 and 1 (GEO-R0, GEO-R1) or by satellite footprints (SatFoot). Crucially, even with partitioning, the bitstring in segmented-BIER remains proportional to the number of user terminals within each partition—partitioning shrinks the address space but does not remove user-count sensitivity—whereas BIER-Star’s cell-encoded header is effectively insensitive to group size.

\begin{figure}[!t]
  \centering
  % Optional: align captions nicely
  % \captionsetup[sub]{justification=centering,singlelinecheck=false}

  \begin{subfigure}[b]{0.7\columnwidth} % Changed width for vertical layout
    \centering
    \includegraphics[width=\linewidth]{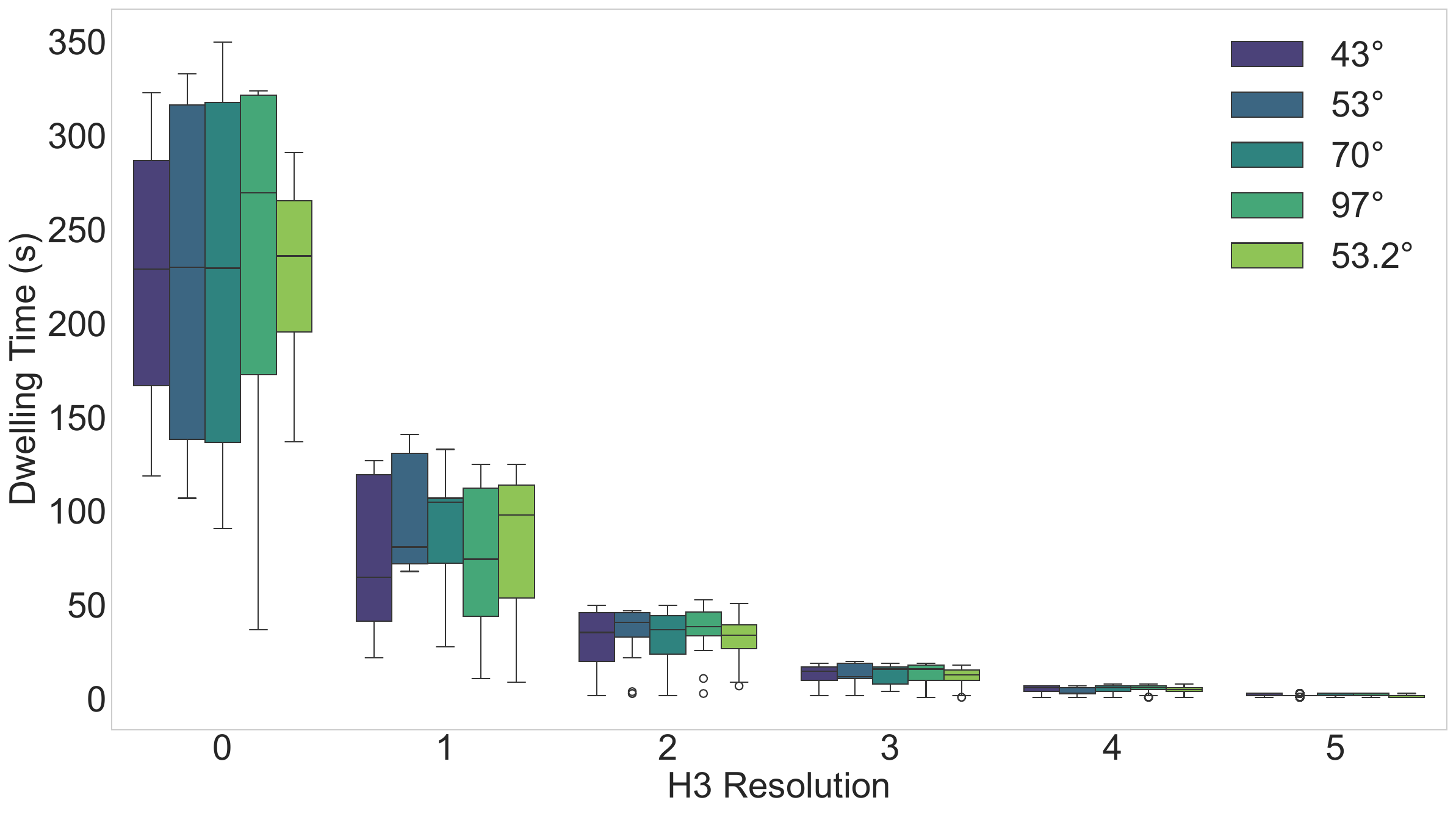}
    \caption{Starlink dwelling time.}
    \label{fig:starlink}
  \end{subfigure}
  
  % A blank line here creates the vertical break
  
  \begin{subfigure}[b]{0.7\columnwidth} % Changed width for vertical layout
    \centering
    \includegraphics[width=\linewidth]{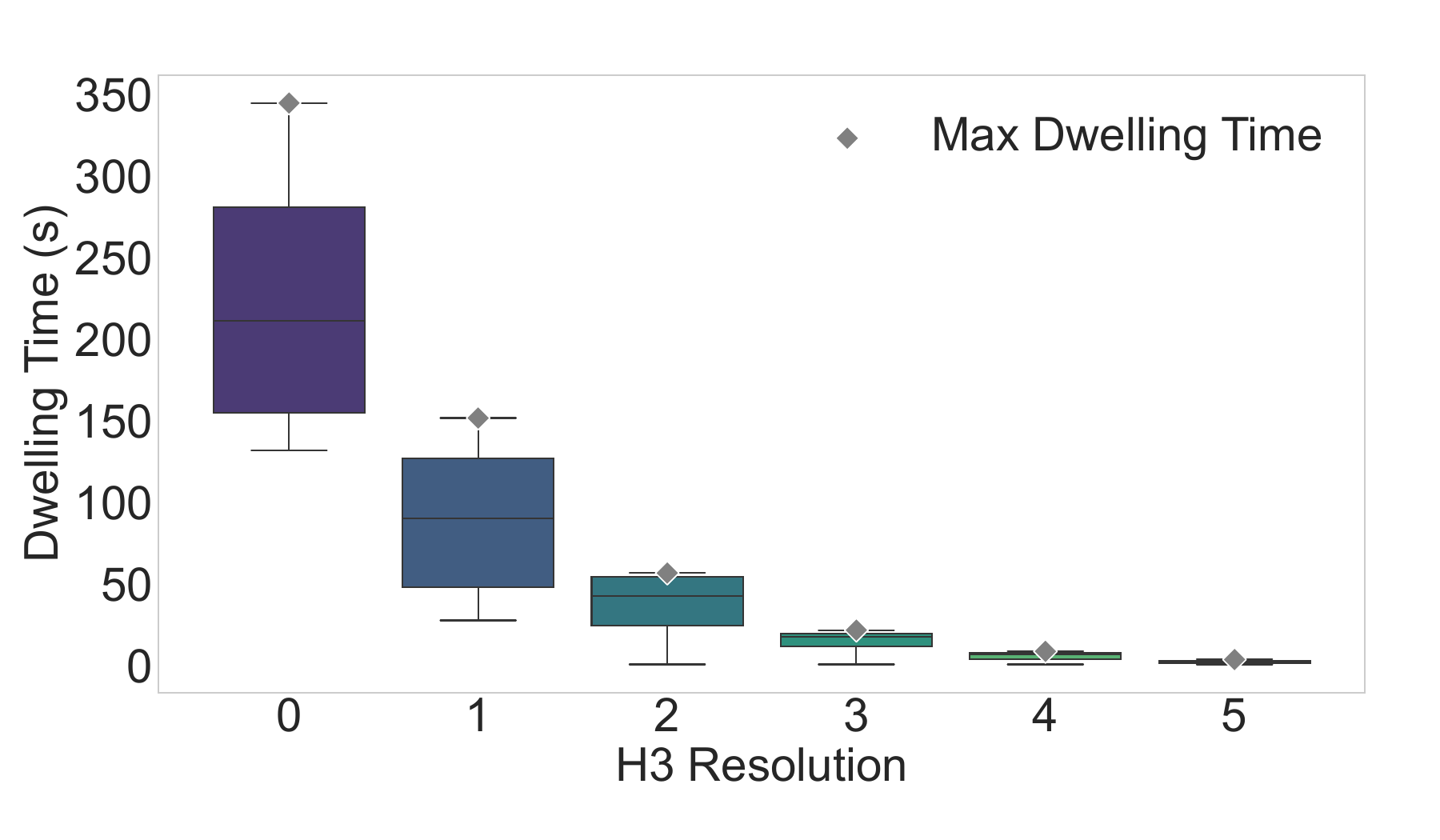}
    \caption{OneWeb dwelling time.}
    \label{fig:oneweb2}
  \end{subfigure}

  \caption{Comparison of different satellite constellations based on orbital inclination and dwelling time distribution.}
  \label{fig:dwellingtime}
\end{figure}

Fig.~\ref{fig:dwellingtime} reports the average dwelling time for the Starlink and OneWeb topologies across multiple inclinations. Longer dwelling times mean a satellite remains within the same H3 cell (at resolution {\it r}) longer, yielding fewer handovers and ISL update events; shorter dwelling times imply faster cell crossings, so the H3 cells along the SPT change more rapidly and the ingress must refresh the encoded path more frequently. As shown, for both Starlink and OneWeb, the average dwelling time at H3 resolutions $r\in\{0,1\}$ exceeds $50\,\mathrm{s}$, whereas at $r\in\{3,4,5\}$ it drops below $15\,\mathrm{s}$ (i.e., less than one handover), indicating that mapping the SPT to H3 cells becomes unstable at finer resolutions.

\par
\begin{figure}[!t]
  \centering
  \begin{subcaptionbox}{Starlink \label{fig:png}}[0.48\columnwidth]
    {\includegraphics[width=\linewidth]{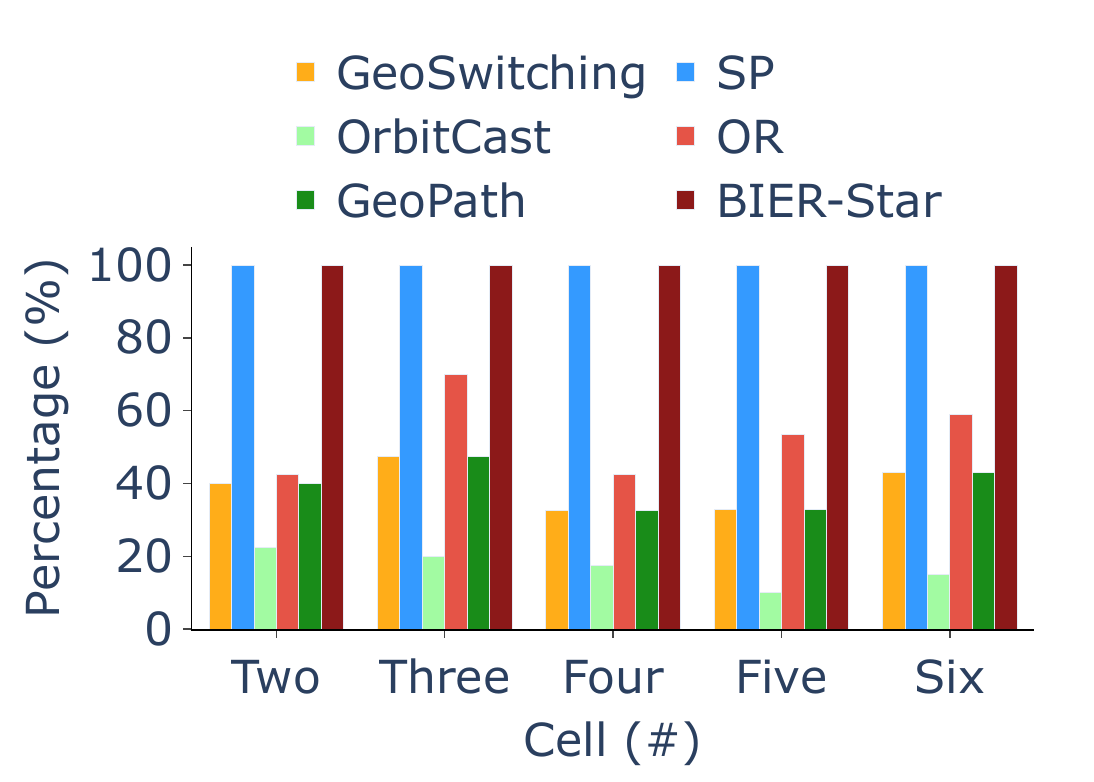}}
  \end{subcaptionbox}\hfill
  \begin{subcaptionbox}{OneWeb \label{fig:eps}}[0.48\columnwidth]
    {\includegraphics[width=\linewidth]{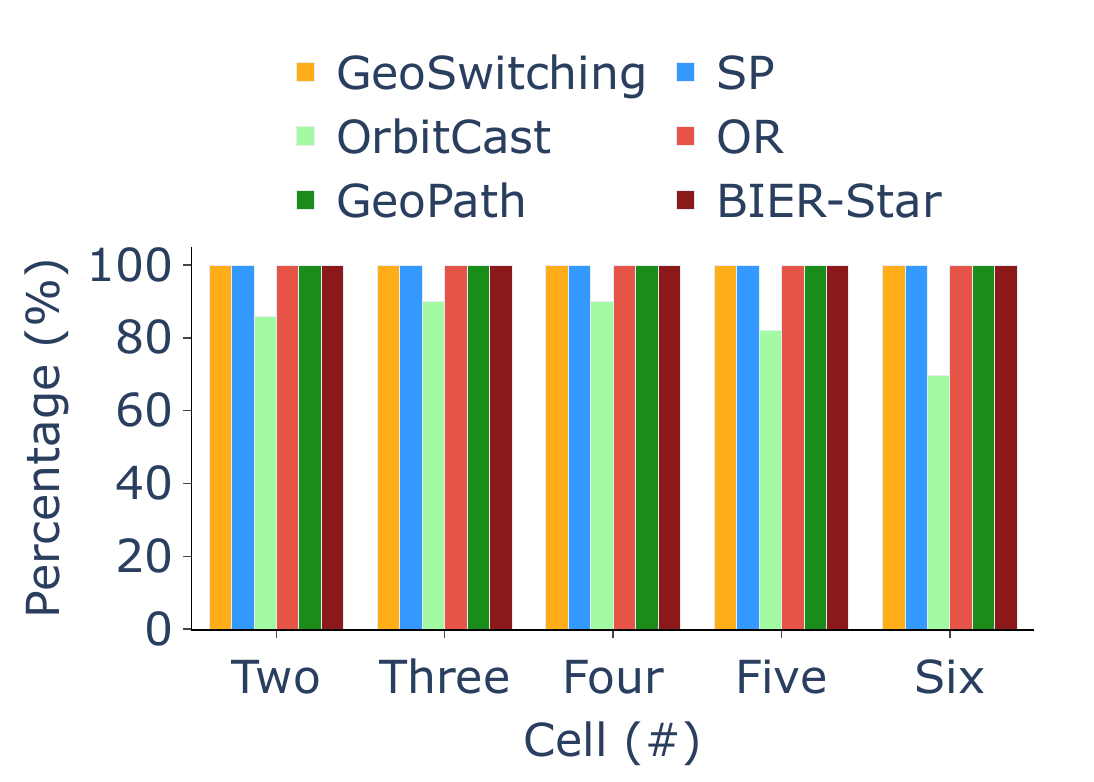}}
  \end{subcaptionbox}
  \caption{Average reach rate of routing schemes on Starlink and OneWeb LEO constellations.}
  \label{fig:Starlink_OneWeb_reachRate}
\end{figure}

 We evaluate BIER-Star’s average reach rate against geographic baselines, OrbitCast \cite{P16}, \cite{P17} (called GeoSwitching), and \cite{P14} (called GeoPath), as well as OR \cite{P15}, extending the originally unicast methods to multicast by routing from the source to each destination and branching at intersections. These geographical methods select the next satellite based on geographical progress and they lack full-path awareness provided by BIER-Star. As show in Fig.~\ref{fig:Starlink_OneWeb_reachRate}, in the Starlink simulation, several greedy geographic methods performed poorly with reach rates as low as 15\% in some scenarios, whereas BIER-Star consistently maintained a reach rate of nearly 100\%. Even in OneWeb’s more regular topology, where most methods achieved near-perfect reach, still the geographical based method(s) cannot find path to some destinations.

\par
\begin{figure}[!t]
  \centering
  \begin{subcaptionbox}{Starlink.\label{fig:png}}[0.49\columnwidth]
    {\includegraphics[width=\linewidth]{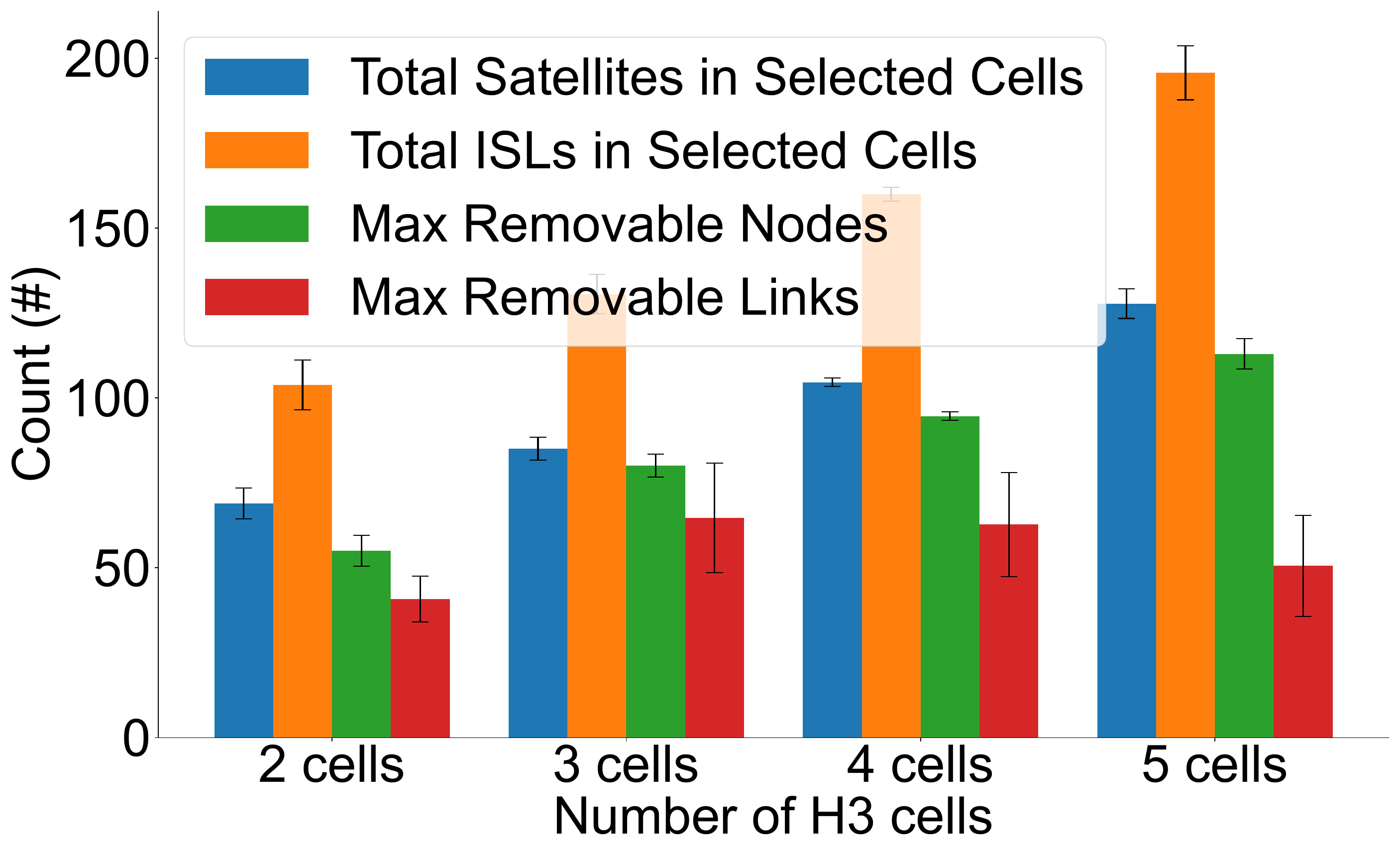}}
  \end{subcaptionbox}\hfill
  \begin{subcaptionbox}{OneWeb. \label{fig:eps}}[0.49\columnwidth]
    {\includegraphics[width=\linewidth]{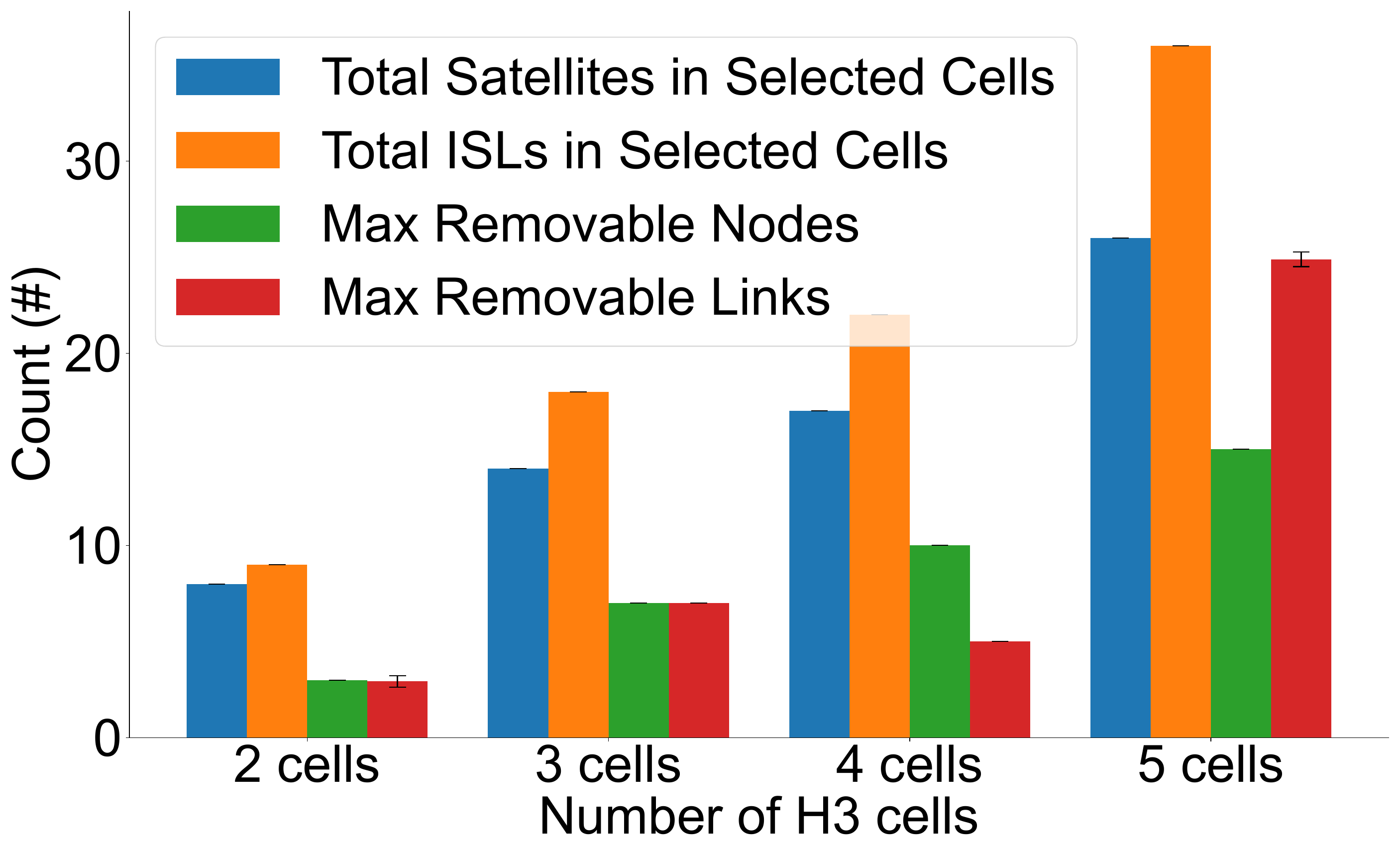}}
  \end{subcaptionbox}
  \caption{Comparative analysis of network resilience in the (a) Starlink and (b) OneWeb constellations.}
  \label{fig:Starlink_OneWeb_resilience}
\end{figure}

 BIER-Star embeds a \emph{complete cell-level route plan} in each packet: from the ingress satellite/gateway to all destinations, the packet carries the sequence (or set) of next H3 cells rather than relying on greedy hops. In mega-constellations where ISL or satellite failures can occur, a forwarder that loses its preferred neighbor can immediately switch to an \emph{alternate satellite in the same H3 cell at resolution $r$}, or fall back to a \emph{parent cell at $r\!-\!1$}, because the packet addresses cells, not nodes; no per-flow state is required. To quantify resilience, we evaluate at H3 resolution~0 the maximum removable nodes and maximum removable links on both Starlink and OneWeb. As shown in Fig.~\ref{fig:Starlink_OneWeb_resilience}), Starlink is more resilient than OneWeb due to its larger number of satellites.

\section{Conclusion}
In large-scale TN-NTN deployments with thousands of satellites and millions of user terminals, stateless multicast schemes that allocate one bit per egress or per link (e.g., BIER/BIER-TE) are fundamentally unscalable because packet headers exceed practical limits. BIER-Star aggregates satellites and user terminals into H3 cells and encodes a compact, cell-level address set in the header. This design decouples header size from group size, eliminates per-flow state at satellites, and aligns forwarding with geography. Our evaluation shows that BIER-Star significantly reduces header length relative to classic and segmented BIER.

 \section*{ Acknowledgment} 
 The work is supported in part by NSERC, CFI and BCKDF.

\bibliographystyle{IEEEtran}
\bibliography{References.bib}{
}

\end{document}